\documentclass[prl,aps,amssymb,shownopacs,twocolumn,superscriptaddress]{revtex4}
\usepackage{amsmath}
\usepackage{amssymb}
\usepackage{amsthm}
\usepackage{amsfonts}
\usepackage{enumerate}
\usepackage{latexsym}
\usepackage[dvips]{graphicx}
\usepackage{epsfig}
\usepackage{subfigure}

\newcommand{\beq}{\begin{equation}}
\newcommand{\eneq}{\end{equation}}

\input{epsf}

\begin{document}

\tolerance 10000

%\draft

\title{Transport Equations and Spin-Charge Propagating Mode \\ in the Two Dimensional Hole Gas}

\author{Taylor L. Hughes}
\affiliation{Department of Physics, Stanford University, Stanford,
CA 94305} \affiliation{IBM-Stanford Spintronic Science and
Applications Center, 650 Harry Rd. San Jose, CA 95120 }
\author{B. Andrei Bernevig}
\affiliation{Department of Physics, Stanford University, Stanford,
CA 94305} \affiliation{IBM-Stanford Spintronic Science and
Applications Center, 650 Harry Rd. San Jose, CA 95120 }
\author{Yaroslaw B. Bazaliy}
\affiliation{IBM-Stanford Spintronic Science and Applications
Center, 650 Harry Rd. San Jose, CA 95120 } \affiliation{IBM Almaden
Research Center, 650 Harry Rd. San Jose, CA 95120}

\begin{abstract}
We find that the spin-charge motion in a strongly confined
two-dimensional hole gas (2DHG) supports a propagating mode of cubic
dispersion apart from the diffusive mode due to momentum scattering.
Propagating modes seem to be a generic property of systems with
spin-orbit coupling. Through a rigorous Keldysh approach, we obtain
the transport equations for the 2DHG, we analyze the behavior of the
hole spin relaxation time, the diffusion coefficients, and the
spin-charge coupled motion.
\end{abstract}
%\pacs{73.43.-f,72.25.Dc,72.25.Hg,85.75.-d}

\maketitle

The correct description of the motion of charge and spin packets in
materials such as semiconductors requires appropriate transport
equations. While transport equations for the Fermi liquid  usually
 describe separate diffusion processes for spin and charge densities
\cite{nozieres1994}, spin-orbit (SO) coupling alters their form
dramatically. For example, the presence of coupling between spin and
charge, a direct result of SO coupling, leads to the possibility of
controlling the spin-orientation by electric fields
\cite{kato2004B}. Another example is the recently proposed
spin-charge mode \cite{bernevig2005} propagating with a velocity
equal to the Rashba SO coupling. This mode has no counterpart in
usual transport and is intimately related to a non-equilibrium
version of the spin-current in the ground-state of the system
\cite{rashba2003}.

In the case of the two-dimensional electron gas (2DEG) with Rashba
spin-orbit interaction, diffusive transport equations  are known
\cite{burkov2003,Mishchenko2004a}. A similar analysis for the
two-dimensional hole gas (2DHG), where the spin-orbit coupling is
much stronger, has so far been lacking. In this letter we obtain the
transport equations for very thin p-doped quantum wells in which
only the heavy-hole band is populated. We start with a general
Keldysh equation for spin-$1/2$ systems with generic SO coupling,
 and then particularize to the case of the heavy hole 2DHG. The
same approach can be applied  other types of SO couplings, such as
the 3D Dresselhauss with little or no modification. The equations
obtained here form the basis for studies of transport and spin-orbit
effects in 2DHG systems. We find that, unlike in the case of 2DEG,
the spin orbit effects do not produce first order derivative terms
in the transport equations, but appear only in second and even third
order derivative terms, which now have to be taken into
consideration. As such, the 2DHG does not exhibit the uniform spin
polarization under an applied electric field, or the spin-galvanic
effect that its 2DEG counterpart does. The diffusion coefficient
becomes a tensor with components for spin and charge which depend on
the spin-orbit coupling. Its spin component turns out to usually be
smaller than the charge component, which might provide an alternate
explanation for the recent experimental
observations\cite{Orenstein2005} attributed to the spin-coulomb
drag. The Dyakonov-Perel relaxation time for the heavy holes' spin
is calculated as well. We discuss the relation of the previously
calculated spin Hall conductivity in the ballistic regime
\cite{schliemann2005,bernevig2004a} and this formalism. In addition,
the transport equations support a spin-charge propagating mode,
analogous to the one predicted in Ref.~\cite{bernevig2005}. However,
its dispersion relation is changed because of the different form of
the SO interaction, and is cubic rather than linear in momentum. The
effects of Coulomb repulsion on the spectrum of the propagating mode
are also accounted for in the present treatment.

The most general single electron Hamiltonian with spin-orbit coupling for spin 1/2 systems is:
\begin{equation}
H = \frac{k^2}{2 m} + \lambda_i(k) \sigma_i, \quad i=1,2,3
\end{equation}
with $\lambda_i(\vec{k})= - \lambda_i(-\vec{k})$  due to time
reversal invariance. The energy eigenvalues are $E_{\pm}(k) = k^2/2m
\pm |\lambda(k)|$ where $|\lambda| = \sqrt{\lambda_i\lambda_i}$. The
retarded and advanced Green's functions
%$G^{R,A}(k, \epsilon) = (\epsilon - H \pm \frac{i}{2\tau})^{-1}$
of the clean 2DHG have the form :
\begin{equation}
G^{R,A}(k, \epsilon) = \frac{1}{2} \sum_{s=\pm 1}
    \frac{1 - s \lambda_i \sigma_i/|\lambda|}
    {\tilde{\epsilon} + s \lambda \pm i/(2\tau)}
\end{equation}
where $\tilde{\epsilon} = \epsilon + k_F^2/(2m) - k^2/(2m)$ and
$\epsilon$ is measured from the Fermi energy $\epsilon_F =
k_F^2/2m$. The density matrix $g = g (k, r, t)$ is a $2 \times 2$
matrix in spin indices, whose integral over the momentum space
yields the reduced density matrix $\rho=\rho(r,t)$ :
\begin{equation}
\rho(r,t)= \int \frac{d^d k}{(2 \pi)^{d+1} \nu} g(k,r,t) \equiv n(r)
+ S_i(r) \sigma_i
\end{equation}
where $n(r,t), S_i(r,t)$ are the charge and spin densities, $d$ is
the space-dimension, and $\nu$ is the $d$-dimensional density of
states.  The charge-spin density $g$ satisfies a Keldysh-type
equation which takes electron scattering into account through an
isotropic momentum relaxation time $\tau$:
\begin{equation}
\label{eq:keldysh_1}
\frac{\partial g}{\partial t} + \frac{g}{\tau} + i \left[H, g
\right] =  \frac{i}{\tau}(G^R \rho - \rho G^A) - \frac{1}{2}
\left\{\frac{\partial H}{\partial k_i}, \frac{\partial g}{\partial
r_i} \right\}
\end{equation}
This quantum Boltzmann equation can be solved iteratively in the
case of small spatial gradients $\nabla g$ of the charge-spin
density \cite{Mishchenko2004a}. Fourier transforming with respect to
time and defining a linear functional $K(g) = \frac{i}{\tau}(G^R
\rho - \rho G^A) -\frac{1}{2} \left\{\frac{\partial H}{\partial
k_i}, \frac{\partial g}{\partial r_i}\right\}$, one obtains an
equation for $g(k,r,\omega)$:
\begin{equation}
g = i \frac{(2 \lambda^2- \Omega^2) K(g) + 2 \lambda_i \lambda_j
\sigma_i K(g) \sigma_j - \Omega \lambda_i [\sigma_i, K(g)]}{\Omega(4
\lambda^2 - \Omega^2)}
\end{equation}
with $\Omega \equiv \omega + i/\tau$. Expanding in gradients $g =
g^{0} + \delta g^{(1)} + \delta g^{(2)} + \ldots$, a perturbative
chain of equations is obtained up to the desired accuracy.
Integrating over the momentum and taking the DC limit $\omega \tau
<< 1$ one obtains the transport equations for $\rho$. In the case of
the Rashba coupling $\lambda_i  = \alpha \epsilon_{ij} k_j$ this was
done in Ref.~\cite{Mishchenko2004a}.

For 2D systems with an isotropic Fermi surface, which includes the
Rashba model, the 2DHG model, and the 2D Dresselhauss model, in the
limit of $ \epsilon_F \tau \ll 1$, $\epsilon_{\alpha} \equiv
|\lambda(k_F)| \ll \epsilon_F$ the continuity equation takes the
form:
\begin{eqnarray}
\label{eq:rho}
\frac{\partial \rho}{\partial t} &=& - \frac{2 \lambda^2}{(4 \lambda^2
    - \Omega^2)\tau} (\rho - \frac{1}{2} \sigma_m \rho \sigma_m)
\\
\nonumber
    &-& i \int \frac{d^2 k}{(2 \pi)^{2} m} \Omega( \delta g^{(1)}
    + \ldots )
\end{eqnarray}
where $\delta g^{(1)}$ contains first order spatial derivatives, and
subsequent terms contain higher order derivatives. Here $m=1,2; \; m
\ne 3$. The first term on the right hand side has zero components in
the charge-sector and represents spin-relaxation with characteristic
time:
\begin{equation}
\label{eq:tau} \frac{1}{\tau_s}  = \frac{1}{\tau} \left[ \frac{2
\zeta^2}
    {4 \zeta^2 +1} \right]
\end{equation}\noindent where $\zeta\equiv \epsilon_\alpha \tau.$
Equation (\ref{eq:rho}) and the relaxation time formula
(\ref{eq:tau}) are valid for any spin-orbit coupling in a 2D system
with an isotropic Fermi surface. The general expressions for higher
order derivative terms tend to be exceedingly long.

We now turn to a particular case: the two-dimensional hole gas. The
valence band of type $\rm{III-V}$ bulk semiconductors is a spin
$3/2$ band, separated into heavy and  light-hole bands which touch
at the $\Gamma$ point. In a quantum well a gap opens at the $\Gamma$
point so that the light hole band is moved below the heavy hole
band. A sufficiently thin well ensures a gap large enough to place
the light-hole band below the Fermi level. Then only the heavy-hole
Hamiltonian should be considered. Asymmetry of the quantum well
introduces a spin-orbit term of the form $\lambda_i \sigma_i$ with:
\begin{equation} \lambda_x =\alpha k_y(3 k_x^2 - k_y^2), \;\;\; \lambda_y =\alpha
k_x(3 k_y^2 - k_x^2)
\end{equation}
\noindent
 and $\lambda_z=0$\cite{winkler2000}. This
model has an isotropic Fermi surface with $|\lambda| = \alpha k^3$.
The limit of validity of our calculations is $\hbar/\tau, \; \alpha
k_F^3 \ll k_F^2/2 m_{hh}$ where $m_{hh}$ is the heavy-hole mass. For
a compact notation, we define the charge-spin four-vector
$N^{\mu}=(n,\vec{S}), \; \mu = 0, 1,2,3$. The generic continuity
equation expanded up to the third order in spatial derivatives has
the form:
\begin{eqnarray}
\nonumber
&& \frac{\partial N^{\mu}}{\partial t} + \frac{ N^{\mu}}{\tau_s} (1-\delta_{\mu,0})=
 C_{i}^{\mu\nu}\partial_i  N^{\nu} + D_{ij}^{\mu\nu}\partial_i \partial_j  N^{\nu}
\\
\label{eq:generic_diffusive}
&& \quad + H_{ijk}^{\mu\nu} \partial_i\partial_j\partial_k  N^{\nu}
\end{eqnarray}
\noindent where $i,j,k = \{x,y\}.$We first concentrate on the
coefficients $C^{\mu\nu}_i$ and $D^{\mu\nu}_{ij}$. In the 2DHG case
we find $C_{i}^{\mu\nu} \equiv 0$ due to the symmetries of the
Hamiltonian. This is in contrast with the linear Rashba model of the
2DEG, where nonzero linear terms were calculated in
Ref.~\cite{bernevig2005}. The diffusion coefficient tensor
$D_{ij}^{\mu\nu}$ has only three distinct elements due to the $x
\leftrightarrows y$ symmetry of the 2DHG Hamiltonian. As we have
already mentioned the general forms for the transport coefficients
are very cumbersome so we assume small SO coupling
($\epsilon_{\alpha} \ll \epsilon_F$) and expand up to linear order
in $\frac{\epsilon_\alpha}{\epsilon_F}$ in all coefficients ($D$'s
and $H$'s) except for the charge diffusion coefficient whose first
correction due to SO coupling is at order
$(\frac{\epsilon_\alpha}{\epsilon_F})^2.$ We find
\begin{eqnarray}
D_{xx}^{00}& =&D_{0} \left( 1+
\frac{9}{8}\left(\frac{\epsilon_\alpha}{\epsilon_F}\right)^2\left(1+\frac{2}{1+4\zeta^2}\right)
\right)
\nonumber\\
D_{xx}^{11}&=&D_0\left(\frac{1+24\zeta^4+
32\zeta^6}{(1+4\zeta^2)^3}\right)
\nonumber\\
D_{xx}^{33}&=&D_{0} \left( \frac{1 - 12\zeta^2}{(1+4
\zeta^2)^3}\right)
\nonumber \\
D_{xy}^{03}&=&\frac{D_0}{\epsilon_F\tau}\left(\frac{3
\zeta^2(4\zeta^2-1)}{(4\zeta^2+1)^2}\right)
\end{eqnarray}
\noindent where $D_0 = \frac{v_F^2 \tau}{2}$, and  with symmetries
$D_{xx}^{\mu\nu}\rightarrow D_{yy}^{\mu\nu},D_{xx}^{11}\rightarrow
D_{yy}^{22}.$ Another notable symmetry is $D_{xy}^{\mu\nu} =
-D_{yx}^{\mu\nu}$ so the off-diagonal terms in the sum
$D_{ij}^{\mu\nu} \partial_i\partial_j N^{\nu}$ cancel. The
expressions above are valid in the limit when the spin-orbit band
splitting and momentum relaxation energy are assumed to be smaller
than the Fermi energy \emph{i.e.} $\epsilon_{\alpha},
\frac{1}{\tau}<<\epsilon_{F}.$ However, the value of $\zeta$ is
unrestricted and we find for some common physical parameters that $0
< x < 1.$ The transport equation above with these diffusion
coefficients represents one of the main results of this paper.

Here we briefly show how an interesting new phenomenon, the spin
Hall effect, emerges from the above equations. Under an applied
electric field along the $\hat{x}$-axis we expect out-of-plane spins
to be transported in the $\hat{y}$
direction\cite{murakami2003,SINOVA2004}. We take into account the
effects of an electric field per the substitution $\vec{\nabla} n
\rightarrow \vec{\nabla} n - e\nu \vec{E}$ \cite{burkov2003}. From
the transport equations above we see that $E^x$  generates:
\begin{equation}
J_{y}^{z}=(D_{yx}^{03}e \nu)E^x\end{equation} \noindent which is in
the preferred form\cite{murakami2003,SINOVA2004} and where the
symmetry $D_{xy}^{\mu\nu}=-D_{yx}^{\mu\nu}$ gives us the
antisymmetric tensor structure $J_{i}^{j}=\sigma_{SH} \epsilon_{ijx}
E^x$. In the low impurity limit \emph{i.e.} as $\tau\rightarrow
\infty$ we find a value of
\begin{equation}
D_{xy}^{03}= \frac{3}{4m}.\end{equation} \noindent This gives us:
$J_{y}^{z}=\frac{3e}{8\pi}E^x.$ When we take into account, like
\cite{schliemann2005} that we are using spin-$1/2$ matrices to
represent a spin-$3/2$ system, we account for an additional factor
of $3$ which yields: \begin{equation}
\sigma_{SH}=\frac{9e}{8\pi}\end{equation}\noindent in agreement with
\cite{bernevig2004a,schliemann2005}. This value for the conductivity
matches the clean-limit and supports the previously known fact that
vertex corrections vanish for this model. \cite{bernevig2004a}

\begin{figure}%[htp]
     \centering
$\begin{array}{cc}
     \subfigure{
          \label{fig2}
          \includegraphics[width=.20\textwidth,height=0.13\textheight]{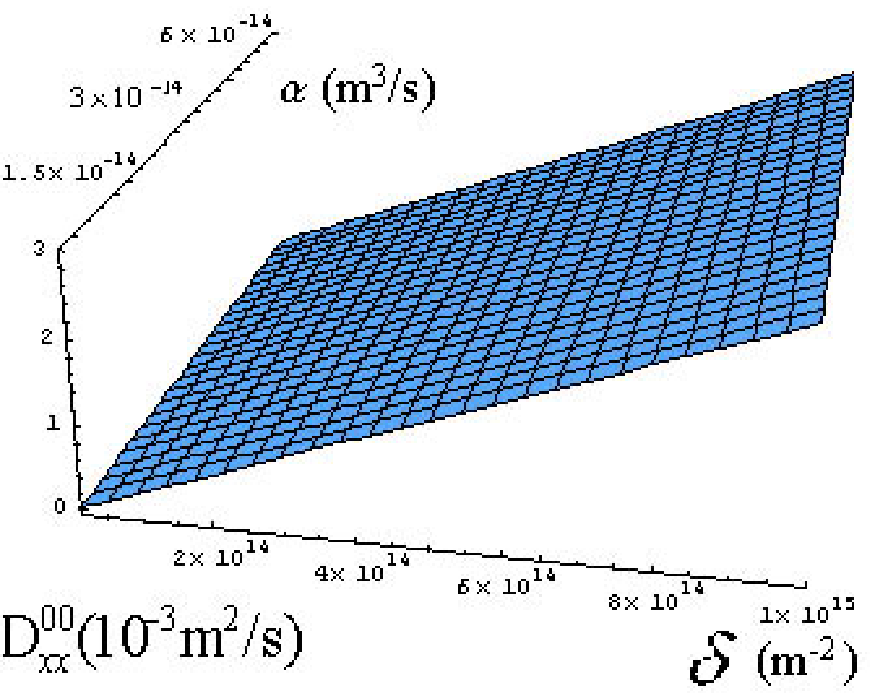}}
     \hspace{.3in}
&\subfigure{
          \label{spinrelaxationtime}
\includegraphics[width=.22\textwidth,height=0.11\textheight]{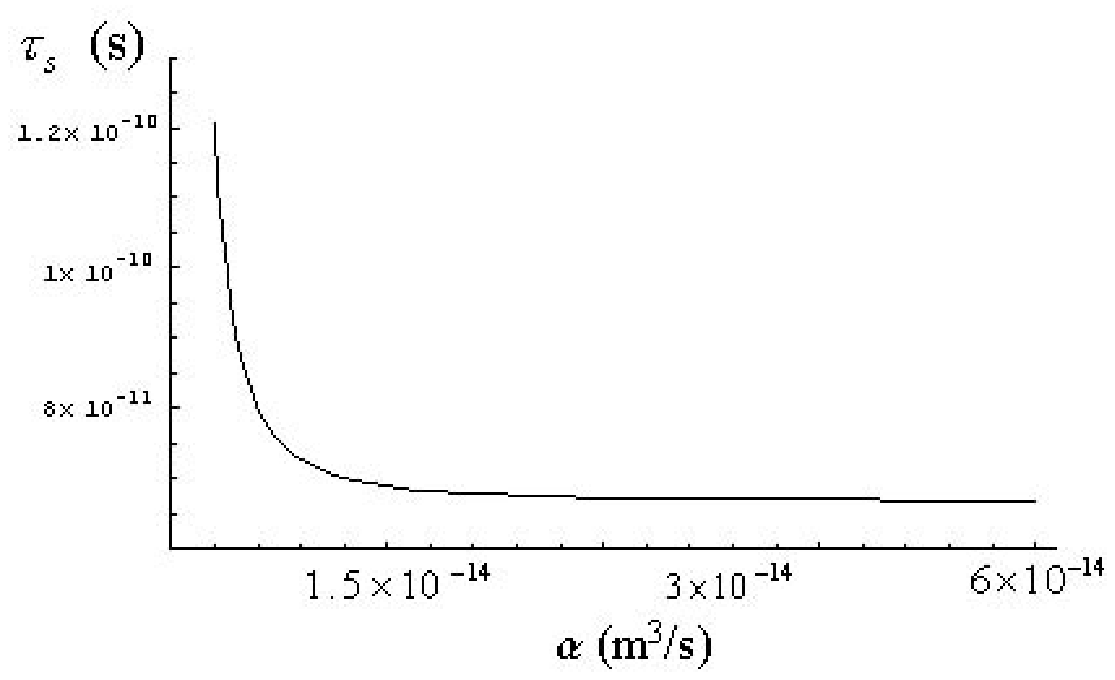}
}\\
     \subfigure{
          \label{fig3}
          \includegraphics[width=.20\textwidth,height=0.13\textheight]{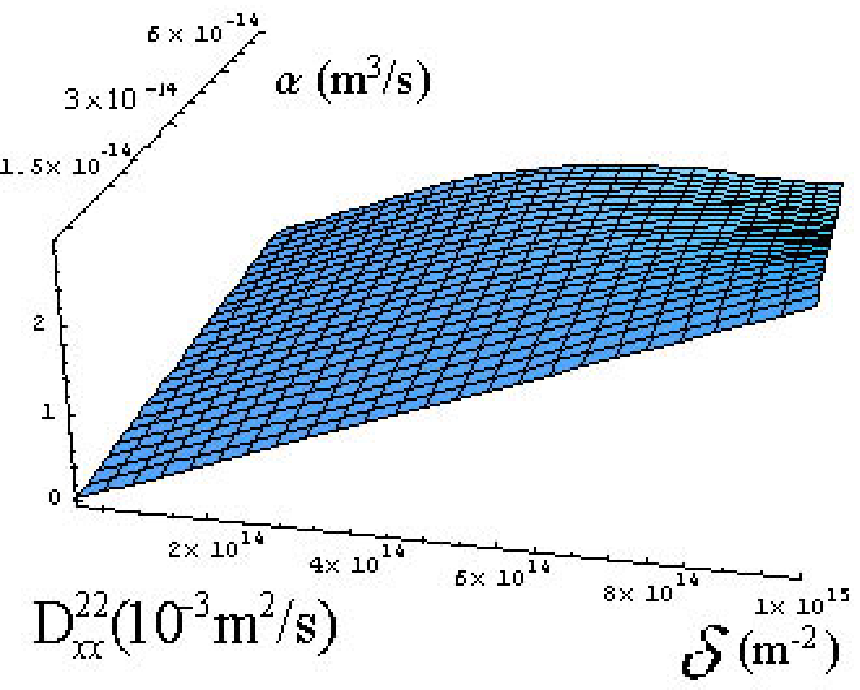}
          }
    \hspace{.05in}
&\subfigure{
          \label{fig4}
\includegraphics[width=.20\textwidth,height=0.13\textheight]{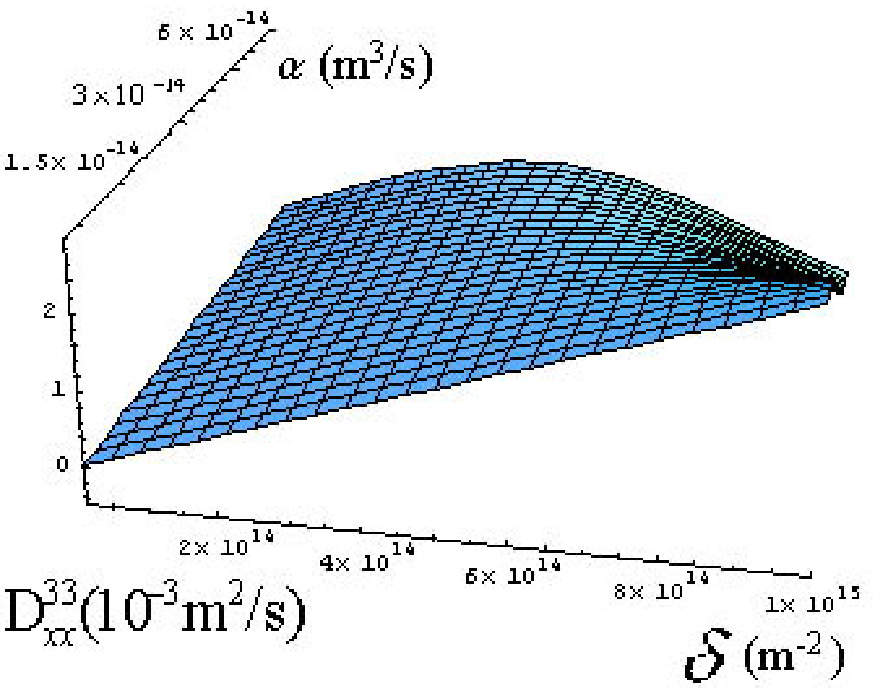}
}
\end{array}$
\caption{(color online) (a)Charge diffusion coefficient(b)Spin
relaxation time plotted vs SO coupling $\alpha$(c) $S^z$ diffusion
coefficient (c) $S^y$ diffusion coefficient. All 3d figures are
plotted vs. SO coupling strength $\alpha$ and hole-doping $\delta.$}
\end{figure}

Nontrivial dynamics, in the absence of applied fields, appears in
the third order derivative terms of the transport equations. The sum
$H_{ijk}^{\mu\nu}
\partial_i\partial_j\partial_k  N^{\nu}$ only depends on
coefficients $H_{xxx}^{\mu\nu}$, $H_{yyy}^{\mu\nu}$, and symmetrized
combinations: ${\tilde H}^{\mu\nu}_{x} =
H^{\mu\nu}_{xxy}+H^{\mu\nu}_{xyx}+H^{\mu\nu}_{yxx}$, and ${\tilde
H}^{\mu\nu}_{y} =
H^{\mu\nu}_{yxy}+H^{\mu\nu}_{yyx}+H^{\mu\nu}_{xyy}$. The number of
distinct coefficients is further reduced to just four by the exact
symmetries of the Hamiltonian: \begin{eqnarray}
H^{02}_{xxx} &=& H^{01}_{yyy}= -\frac{1}{3}{\tilde H}^{02}_{y}= -\frac{1}{3} {\tilde H}^{01}_{x}\nonumber \\
H^{20}_{xxx} &=& H^{10}_{yyy}=-\frac{1}{3} {\tilde H}^{20}_{y} = -\frac{1}{3} {\tilde H}^{10}_{x}\nonumber \\
H^{13}_{xxx} &=& -H^{23}_{yyy} = -\frac{1}{3} {\tilde H}^{13}_{y} =\frac{1}{3} {\tilde H}^{23}_{x} \nonumber \\
H^{31}_{xxx} &=& -H^{32}_{yyy} = -\frac{1}{3} {\tilde
H}^{31}_{y}=\frac{1}{3} {\tilde H}^{32}_{x}.\end{eqnarray} \noindent
For the third-order coefficients we find:
\begin{eqnarray}
H^{02}_{xxx}&=&H_0\left(\frac{3\zeta(3+16\zeta^2+72\zeta^4+24\zeta^6)}{2(1+4\zeta^2)^3}\right)
\nonumber\\
H^{13}_{xxx} &=&H_0\left(\frac{(4\epsilon_F\tau
\zeta)(4\zeta^2-1)}{(1+4\zeta^2)^4}\right)
\end{eqnarray} \noindent where $H_0= \frac{v_F \tau}{2 m}$ and up to this order we have
the additional symmetries: $H_{xxx}^{02}=H_{xxx}^{20}$ and
$H_{xxx}^{13}=-H_{xxx}^{31}.$

\begin{figure}%[htp]
$\begin{array}{cc}
     \subfigure{
          \label{fig5}
          \includegraphics[width=.20\textwidth,height=0.13\textheight]{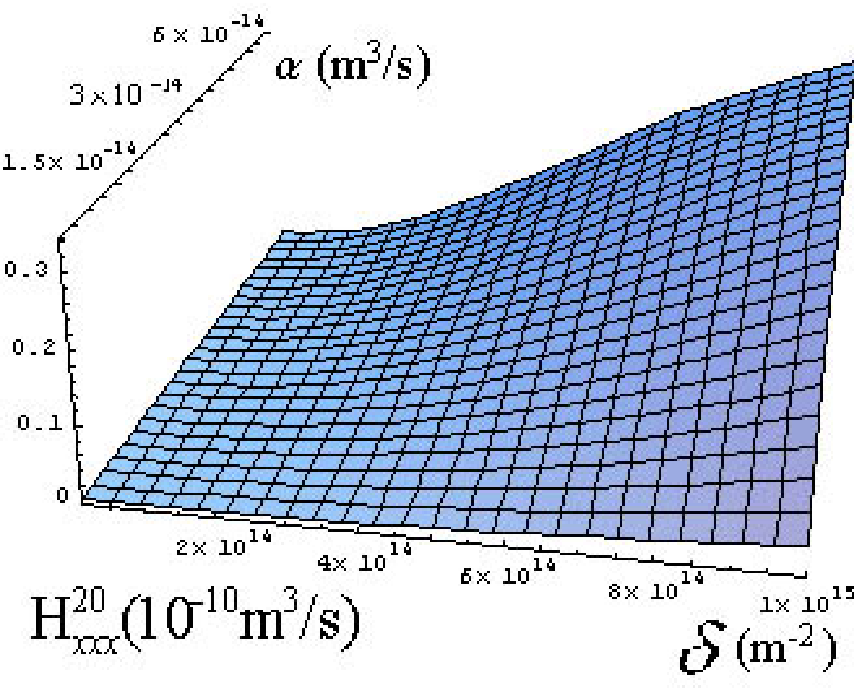}
          }
    \hspace{.05in}
&\subfigure{
          \label{fig6}
\includegraphics[width=.20\textwidth,height=0.13\textheight]{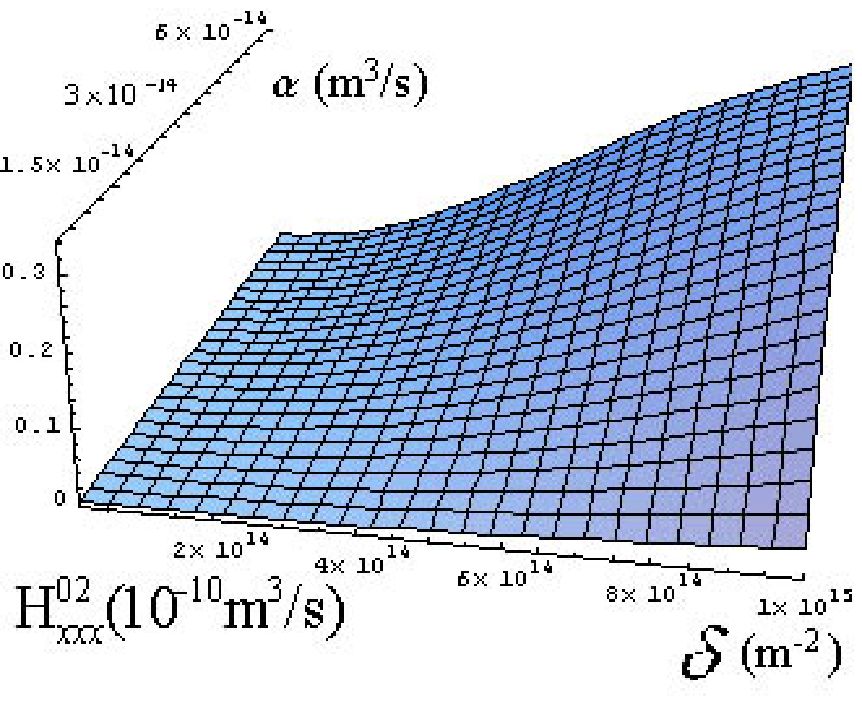}}
\\
   \subfigure{
          \label{fig7}
          \includegraphics[width=.20\textwidth,height=0.13\textheight]{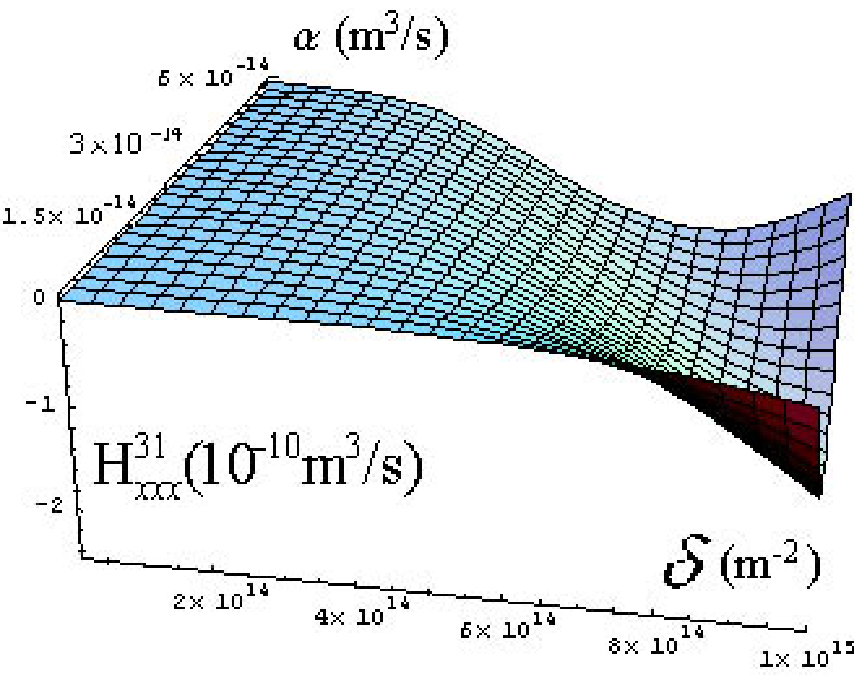}
          }
    \hspace{.05in}
&\subfigure{
          \label{fig8}
\includegraphics[width=.20\textwidth,height=0.13\textheight]{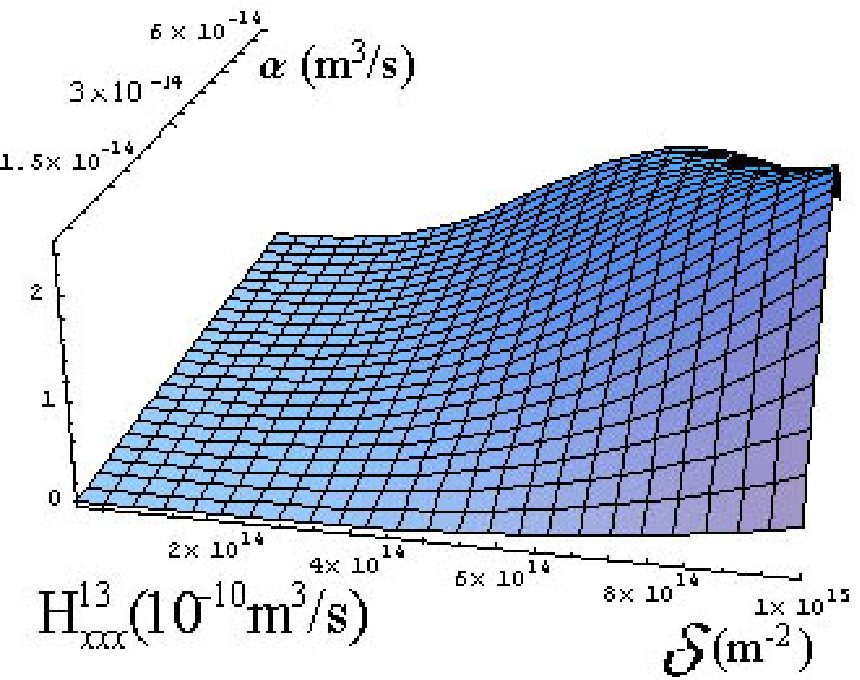}
}
\end{array}$
\caption{(color online) Third order coefficients plotted vs. SO
coupling strength $\alpha$ and hole-doping $\delta.$ }
  \end{figure}

It was found in Ref.\cite{bernevig2005} that in 2DEG systems the SO
interaction can produce a mode of coupled spin and charge
oscillations (fractionalized spin packets mode, or spin-galvanic
mode) which can be weakly damped even in the diffusive regime. It
was later found that the Coulomb interactions of the charge
component do not suppress this mode \cite{bazaliy2005}. Here we
study the analogue of such mode in the 2DHG system using the ansatz
 $\delta n(x,t) = \delta n \exp[iqx - i\omega t]$, $S_y(x,t) =
S_y \exp[iqx - i\omega t]$, $S_x, S_z = 0$ to find a solution of
Eqs.~(\ref{eq:generic_diffusive}) in the absence of external fields.
Following the procedure of Ref.~\cite{bazaliy2005} to account for
Coulomb interactions, we obtain the spectrum of the mode
\begin{equation}
\label{mode_dispersion}
\omega_{\pm}(q) = \frac{-i(\phi_1 + \phi_2) \pm \sqrt{4 \gamma^2 \phi_3 - (\phi_1 - \phi_2)^2}
}{2\tau_s}
\end{equation}
with
$$
\phi_1 = \frac{D_{xx}^{00}}{D_0} \kappa^2 R,
\quad  \phi_2 = \frac{D_{xx}^{22}}{D_0} \kappa^2 + 1 \ ,
\quad  \phi_3 = \kappa^6 R \ ,
$$
where $\kappa = q l_s$, $l_s = \sqrt{D_0 \tau_s}$ is the spin-diffusion length,
$R = 1 + 2\pi\kappa_B/\kappa$ and  $\kappa_B = \nu e^2 l_s$ characterize the
strength of the Coulomb interaction, and $\gamma = H^{02}_{xxx} \tau_s/l_s^3$. As
in the 2DEG case, the shape of the spectrum is controlled by two dimensionless
parameters $\gamma$ and $\kappa_B$. The estimate for $\gamma \sim \epsilon_{\alpha}\tau_p/k_F l_s$ can
 be compared with the 2DEG estimate $\gamma_{2DEG} \sim \epsilon^{2DEG}_{\alpha}\tau_s/k_F l_s$.
 One observes that the larger SO splitting $\epsilon_{\alpha}$ associated with
  2DHG systems can be compensated by the small factor $\tau_p/\tau_s$.
  The Coulomb interaction parameter is large, $\kappa_B \sim 10^3$ for $l_s = 100 nm$.

\begin{figure}[t]
    \resizebox{.45\textwidth}{!}{\includegraphics{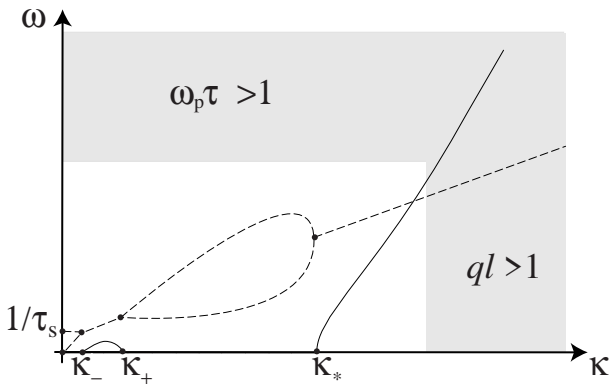}}
\caption{Real (solid line) and imaginary (dashed lines) parts of the mode
frequencies $\omega_{\pm}(k)$ in an infinite 2DHG structure with
 Coulomb repulsion. Two windows of $Re(\omega) \neq 0$ exist.
 The grey areas are outside the limits of the diffusive approximation.}
 \label{fig:spectrum}
\end{figure}

A sketch of the spectrum is shown in Fig.~\ref{fig:spectrum}. As in
the case of 2DEG, the modes are purely dissipative at $\kappa = 0$,
representing the charge diffusion and spin relaxation. Spin and
charge motions get coupled at $\kappa > 0$  and two intervals with
$Re(\omega) \neq 0$ emerge. As in the case of 2DEG, the position of
these intervals is very sensitive to the strength of the Coulomb
interaction $\kappa_B$. However, in addition to that, the
$Re(\omega)/Im(\omega)$ ratio is also affected by the Coulomb
interaction in the 2DHG case. Our calculation shows that the mode is
always highly damped in the first interval. The mode can be weakly
damped with $Re(\omega)/Im(\omega) \gg 1$ in the second window
$\kappa > \kappa_*$, however $\kappa_* \sim
(\kappa_B/\gamma^2)^{1/3}$ is usually too close to the boundary of
the validity of the diffusive approximation which requires $\kappa <
\sqrt{\tau_s/\tau_p}$. We conclude that observing the fractionalized
spin packets propagating mode in the 2DHG requires more fine tuning
of parameters than in the 2DEG case. However, the presence of the
weakly damped solution at the boundary between the diffusive and
ballistic regimes suggests that a search for a ballistic counterpart
of the mode has to be performed.

In conclusion, in the present letter we derived the diffusive
transport equations for the 2DHG system. Our calculation shows that
the components of the diffusion coefficient tensor depend on the
magnitude of the SO coupling and that the spin diffusion is
generally smaller than the charge diffusion, thus offering an
alternate explanation to spin-coulomb drag for this already
experimentally observed \cite{Orenstein2005} phenomenon. We find
that, in contrast to the 2DEG case, the spin-charge coupling enters
into the transport equations starting with the second order
derivative terms. Our diffusive equations reproduce the value of the
spin Hall coefficient, previously calculated for this system in the
ballistic regime \cite{schliemann2005, bernevig2004a}, thus
supporting the conclusion about the absence of vertex corrections in
this 2DHG system. The fractionalized-spin packets propagating mode
in the 2DHG is also studied. It is found that the Coulomb
interaction suppresses this mode more than in the case of electron
gas. The spectrum of the mode suggests that its counterpart may
exist for 2DHG systems in the ballistic regime.

The authors a grateful to A. MacDonald, S. Parkin, D. Arovas and H.
Manoharan for useful discussions. B.A.B acknowledges support from
the SGF. T.L.H. acknowledges support from NSF. This work is
supported by the NSF under grant numbers DMR-0342832 and the US
department of Energy, Office of Basic Energy Sciences under contract
DE-AC03-76SF00515.

\bibliography{spinhall,extra}
\end{document}